\documentclass[pra,twocolumn,notitlepage,superscriptaddress,longbibliography]{revtex4-2}
\usepackage{amsmath}
\usepackage{amssymb}

\usepackage[unicode=true,colorlinks=true,citecolor=blue,urlcolor=blue]{hyperref}

\usepackage{bm}
\usepackage{epsfig}

\usepackage[normalem]{ulem}

\renewcommand {\phi}{{\varphi}}
\newcommand {\rmi}{{\rm i}}

\newcommand {\e}{{\rm e}}
\newcommand {\eps}{\varepsilon}

\begin{document}
\title{
{Two-photon pulse scattering spectroscopy for  arrays of two-level atoms, coupled to the waveguide}
}

\author{Ekaterina Vlasiuk}
\affiliation{Department of Physics, University of Basel, Klingelbergstrasse 82, CH-4056 Basel, Switzerland}

\author{Alexander V. Poshakinskiy}
\affiliation{Ioffe Institute, St. Petersburg 194021, Russia}

\author{Alexander N. Poddubny}
\email{alexander.poddubnyy@weizmann.ac.il}
\affiliation{Department of Physics of Complex Systems, Weizmann Institute of Science, Rehovot 7610001, Israel}

\begin{abstract}
We have theoretically studied the scattering of two-photon pulses from a spatially-separated array of two-level atoms coupled to the waveguide. 
A general analytical expression for the scattered pulse has been obtained. The contributions  of various single-eigenstate 
and double-excited eigenstates of the array have been analyzed.
We have also calculated the dependence of the time incident photons are stored in the array on its period and the number of atoms. The largest storage times correspond to the structures with the anti-Bragg period, equal to the quarter of the wavelength of light at the atom resonance frequency $\lambda/4$.
\end{abstract}
\date{\today}
\maketitle

\section{Introduction}
Waveguide quantum electrodynamics, focused on light-matter interactions in arrays of natural or artificial atoms that are coupled to the waveguide, is a rapidly developing field of quantum optics~\cite{Sheremet,Roy2017}. Multiple experimental platforms to study fundamental physics models and engineer atom-photon interactions have now emerged.
Practical applications such as detection, processing~\cite{Prasad2020} and generation~\cite{Kannan2023} of quantum photon states  would ultimately require devices operating in the pulsed regime. Hence, there is a fundamental need to understand the time-dependent atom-photon interaction in this system.

In fact, a general scattering matrix consideration of the time-dependent photon scattering has already been performed in the first works in the field~\cite{Yudson1984,Yudson2008,Shen2007}. More recent studies have also emphasized the importance of the formation of bound photon states in the time-dependent photon transmission~\cite{Mahmoodian2020,Chen2020,Calajo2022}, complex correlated multi-photon states~\cite{Iversen2022}  as well as have taken into account the time-entangled nature of the photon pulse~\cite{Kiilerich2020,Yang2022}. However, there is one interesting aspect of the time-dependent photon transmission through the arrays, that has so far not been analyzed in detail to the best of our knowledge. That is the role of different collective states of the array. In particular, it is now well known that spatially separated arrays have a complicated structure of collective single- and double-excited states~\cite{Molmer2019,Ke2019,Poshakinskiy2020,Poshakinskiy2021dimer}. These states can be distinguished by their spontaneous decay rate, which can be either enhanced (for superradiant states) or suppressed (for subradiant states) due to the interference between the photons emitted from different atoms. It is then natural to examine the signatures of these states in the photon time dynamics. We have recently analyzed the case of continuous wave excitation in detail in Refs.~\cite{PoshakinskiyBorrmann,NewKornovan}. Here we focus on the case of pulse excitation through the waveguide, see Fig.~\ref{fig:1}. We show that the measurement of the time-dependent joint detection probability of transmitted photons provides additional information about the double-excited collective modes that is not captured by the continuous excitation scheme.

We hope that this work will be useful both as a first step for potential future analysis of a multi-photon pulsed regime and as a helpful tool for analysis of ongoing and coming experiments~\cite{Brehm2022,LeJeannic2022}.

The rest of the manuscript is organized as follows. Section~\ref{sec:model} outlines our 
model and the calculation approach. Next, we present the results for the two-photon wavefunction in Sec.~\ref{sec:two}. Our main results are summarized in Sec.~\ref{sec:summary}. Appendix~\ref{sec:Appendix} is reserved for auxiliary theoretical details.

\section{Model and calculation approach}\label{sec:model}
We consider a basic WQED setup with  $N$  two-level qubits, periodically spaced near a waveguide and interacting via a waveguide mode. 
\begin{figure}[t] 
\includegraphics[width=0.45\textwidth]{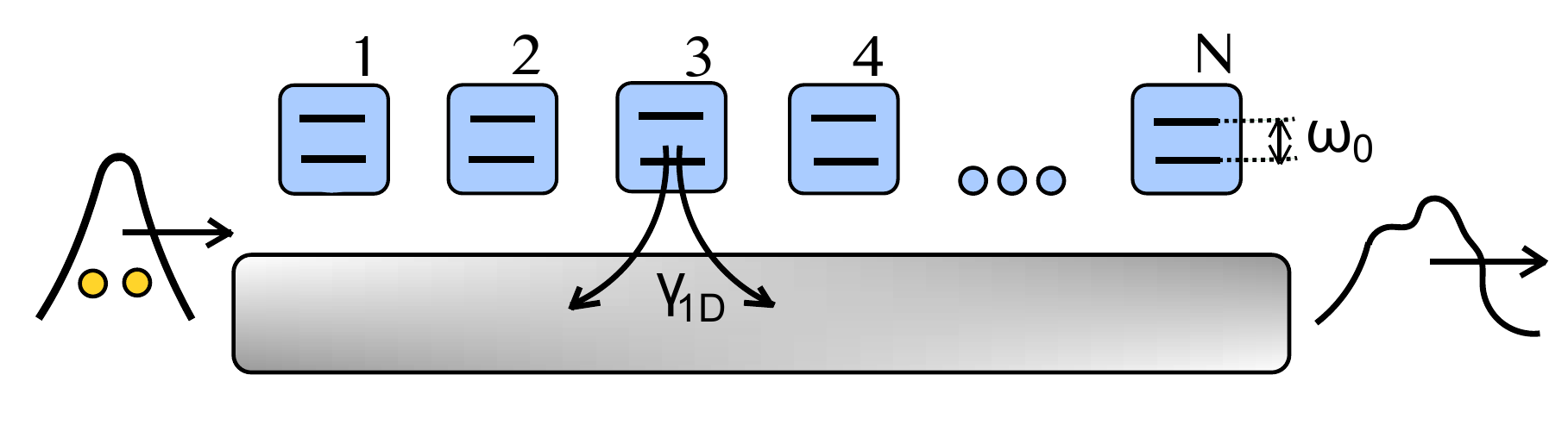}
\caption{Schematics of a two-photon pulse propagating through an array of qubits coupled to a waveguide. Here, $\omega_0$ is a resonant frequency of qubits, $\gamma_{\text{1D}}$ is a spontaneous emission rate into the guided mode.
   } \label{fig:1}
\end{figure}
The system is schematically shown in Fig.~\ref{fig:1} and can be described by the following effective Hamiltonian \cite{Caneva2015,Ke2019,Sheremet},
\begin{equation}
H=-\rmi \gamma_{\rm 1D}\sum_{n,m=1}^N\sigma_n^\dag \sigma^{\vphantom{\dag}}_m \e^{\rmi (\omega_0/c)|z_m-z_n|}\:.
\end{equation}
This Hamiltonian assumes usual Markovian and rotating-wave approximations. The energy is counted from the atomic resonance $\hbar\omega_0$, $c$ is the speed of light, and $z_m$ are the qubit coordinates along the waveguide. For a  periodic array, where $z_{m+1}-z_m=d$, the period can be conveniently  characterized just by a single dimensionless parameter, the phase 
$\varphi=\omega_0d/c\equiv 2\pi d/\lambda_0$  gained by light between two neighbouring two-level atoms with the distance $d$.  The raising  operators $\sigma_m^\dag$ obey the usual spin-1/2 operators algebra:
$\sigma_m^2=0$, $\sigma_m^{\vphantom{\dag}}\sigma_m^\dag+\sigma_m^\dag \sigma^{\vphantom{\dag}}_m=1$, $[\sigma_m,\sigma_n]=0$ for $m \ne n$. The parameter $\gamma_{\rm 1D}$ is the radiative decay rate of  a single atom into the waveguide. It makes  the  effective Hamiltonian  non-Hermitian.

We are interested in the scattering of a general two-photon state, characterized by a 
time-dependent wave function $\psi_{t_1,t_2}^{\text{in}}$  from such a setup. 
To this end we use the known general technique \cite{Laakso2014,Fang2014,Poshakinskiy2016,Ke2019} to calculate the two-photon scattering matrix in the frequency domain $S(\omega'_{1},\omega'_{2}\leftarrow\omega_{1},\omega_{2})$. We start with the Fourier transform of the input state:
\begin{equation}\label{eq:Fourier}
    \psi_{\omega_1,\omega_2}^{\text{in}}=\iint dt_1dt_2e^{\mathrm{i}\omega_1t_1}e^{\mathrm{i}\omega_2t_2}\psi_{t_1,t_2}^{\text{in}}\:.
\end{equation}
The output state is then given by
\begin{equation}
\begin{gathered}    \psi_{\omega_1',\omega_2'}^{\text{out}}=\frac12\iint \frac{d\omega_1d\omega_2}{(2\pi)^2} S(\omega'_{1},\omega'_{2}\leftarrow\omega_{1},\omega_{2})
\psi_{\omega_1,\omega_2}^{\text{in}}\:,
\end{gathered}  
\end{equation}
and then we perform the inverse Fourier transform
\begin{equation}\label{eq:psit}
    \psi_{t_1,t_2}^{\text{out}}=\iint \frac{d\omega_1'd\omega_2'}{(2\pi)^2}e^{-\mathrm{i}\omega_1't_1}e^{-\mathrm{i}\omega_2't_2}\psi_{\omega_1',\omega_2'}^{\text{out}}\:.
\end{equation}
The detailed derivation of the scattering matrix for an arbitrary number and positions of the qubits, mostly following Refs.~\cite{Fang2014,Ke2019},  can be found e.g. in Ref.~\cite{Sheremet}. Here we just recall the answer:
\begin{align}  \label{eq:S}  &S(\omega'_{1},\omega'_{2}\leftarrow\omega_{1},\omega_{2})=\\&(2\pi)^2t_{\omega_1}t_{\omega_2}\nonumber[\delta(\omega_1-\omega_1')\delta(\omega_2-\omega_2')+\delta(\omega_1-\omega_2')\delta(\omega_2-\omega_1')]\nonumber\\
    &+2\gamma_{\text{1D}}^2\sum_{m,n=1}^Ns_m^-(\omega'_{1})s_m^-(\omega'_{2})[\Sigma^{-1}]_{mn}s_n^+(\omega_{1})s_n^+(\omega_{2})\nonumber\\
&\times 2\pi\delta(\omega_1+\omega_2-\omega_1'-\omega_2')\nonumber\:,
\end{align}
where
\begin{equation}\label{eq:Sigma}
    \Sigma_{mn}(\varepsilon)=\int G_{mn}(\omega)G_{mn}(2\varepsilon-\omega)\frac{d\omega}{2\pi}
\end{equation}
is the  self-energy matrix for double-excited states with $G_{ij}$ being the Green function for a single excitation of the array,  given by the inverse of the following matrix:
\begin{multline}\label{eq:G}
    [G^{-1}(\omega)]_{mn}\equiv \omega \delta_{mn}-H_{mn}\\
    =(\omega-\omega_0)\delta_{mn}+\mathrm{i}\gamma_{\text{1D}}e^{\mathrm{i}(\omega_0/c)|z_m-z_n|}\:.
\end{multline}
The coefficients 
\begin{equation}
    s_m^{\pm}(\omega)=\sum_n G_{mn}e^{{\pm}\mathrm{i}(\omega_0/c)z_n}
\end{equation} 
describe the coupling of the array with the incoming and outgoing plane waves.
The first term in the scattering matrix \eqref{eq:S} accounts for the independent photon transmission with the transmission coefficients given by
\begin{equation}
    t_\omega=1-\rmi\gamma_{\rm 1D}\sum_{mn} G_{mn}e^{\mathrm{i}(\omega_0/c)(z_n-z_m)}\:.
\end{equation} 
The second term in Eq.~\eqref{eq:S} accounts for the interaction between the photons induced by the array.

For a general pulse shape, the integration over time and frequency can be performed only numerically and is rather tedious. However, it is greatly simplified for a short pulse when the input state  $\psi_{t_1,t_2}^{\rm in}$ can be approximated by a product of two $\delta$-functions, 
\begin{equation}
 \psi_{t_1,t_2}^{in}=\delta(t_1)\delta(t_2)\label{eq:psiin}\:.
\end{equation}
Physically, this means that the pulse duration is significantly shorter than the inverse rate decay of the fastest eigenmode of the system, that is on the order of $1/(N\gamma_{\rm 1D})$. From now on we restrict ourselves to such a case. The calculation procedure is detailed in Appendix~\ref{sec:Appendix}.

\section{Transmitted pulse}\label{sec:two}
We start this section by analyzing in detail the wave function for the pulse, transmitted through the subwavelength array of a given length $N=4$. Next, in Sec.~\ref{sec:N}, we examine the dependence of the effective time it takes the system to scatter photons on the array length $N$.
\subsection{Single- and double-excited states in the transmitted pulse}\label{sec:wavefunction}

\begin{figure*}[t] 
\includegraphics[width=0.9\textwidth]{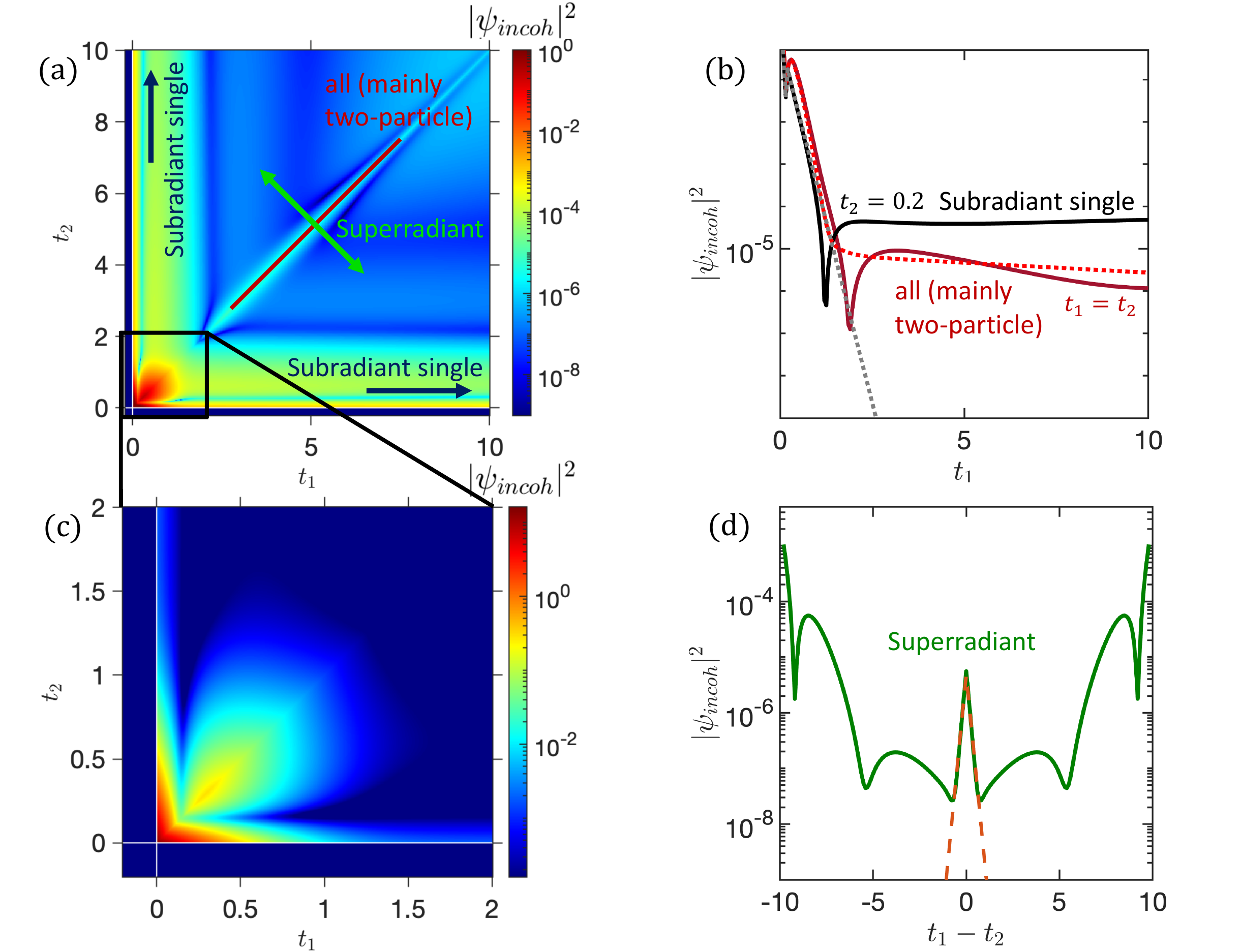}
\caption{Incoherent part of the transmitted pulse for the incident delta-pulse in the time domain. Parameters of the system: $N=4$, $\varphi=0.1$. Times are normalized by $1/\gamma_{\rm 1D}$. (a) Schematics of the various types of photon states in the transmitted pulse. (b) Dark red solid curve: probability distribution to detect two photons at the same moment in time $t_1$. Black solid curve: probability distribution to detect one photon at the time $t_1$ while the second photon is detected at time $t_2=0.2$. Dotted curves have been calculated with all two-particle modes and only the superradiant single-particle mode; the red one -- for the times $t_1=t_2$, the gray one -- for the fixed $t_2=0.2$ (c) Schematics of the various types of photon states in the transmitted pulse. (d) Dependence of the probability of the two photons detection on the time difference $t_1-t_2$ for fixed $t_1+t_2=10$. The solid green curve has been calculated exactly, and the dashed orange curve includes only the superradiant single-excited state and all double-excited ones.} \label{fig:2}
\end{figure*}
Figure~\ref{fig:2} shows the incoherent part of the two-photon wavefunction given by Eq.~\eqref{PsiIncoh} calculated  under the incidence of the two-photon
$\delta$-pulse Eq.~\eqref{eq:psiin}. The incoherent part has quite a complicated time dependence with several distinct time scales.  The shortest time scale $t\sim 1/(N\gamma_{\rm 1D})$ corresponds to the superradiant state where the constructive interference enhances the emission rate. The longest time scale is on the order of 
$N^3/(\varphi^2 \gamma_{\rm 1D})$~\cite{Molmer2019} and corresponds to the excitation of subradiant states.
In order to represent different timescales better, we show the wavefunction at large and short times in Fig.~\ref{fig:2}(a) and Fig.~\ref{fig:2}(c) separately.

\begin{figure}[t] 
\includegraphics[width=0.35\textwidth]{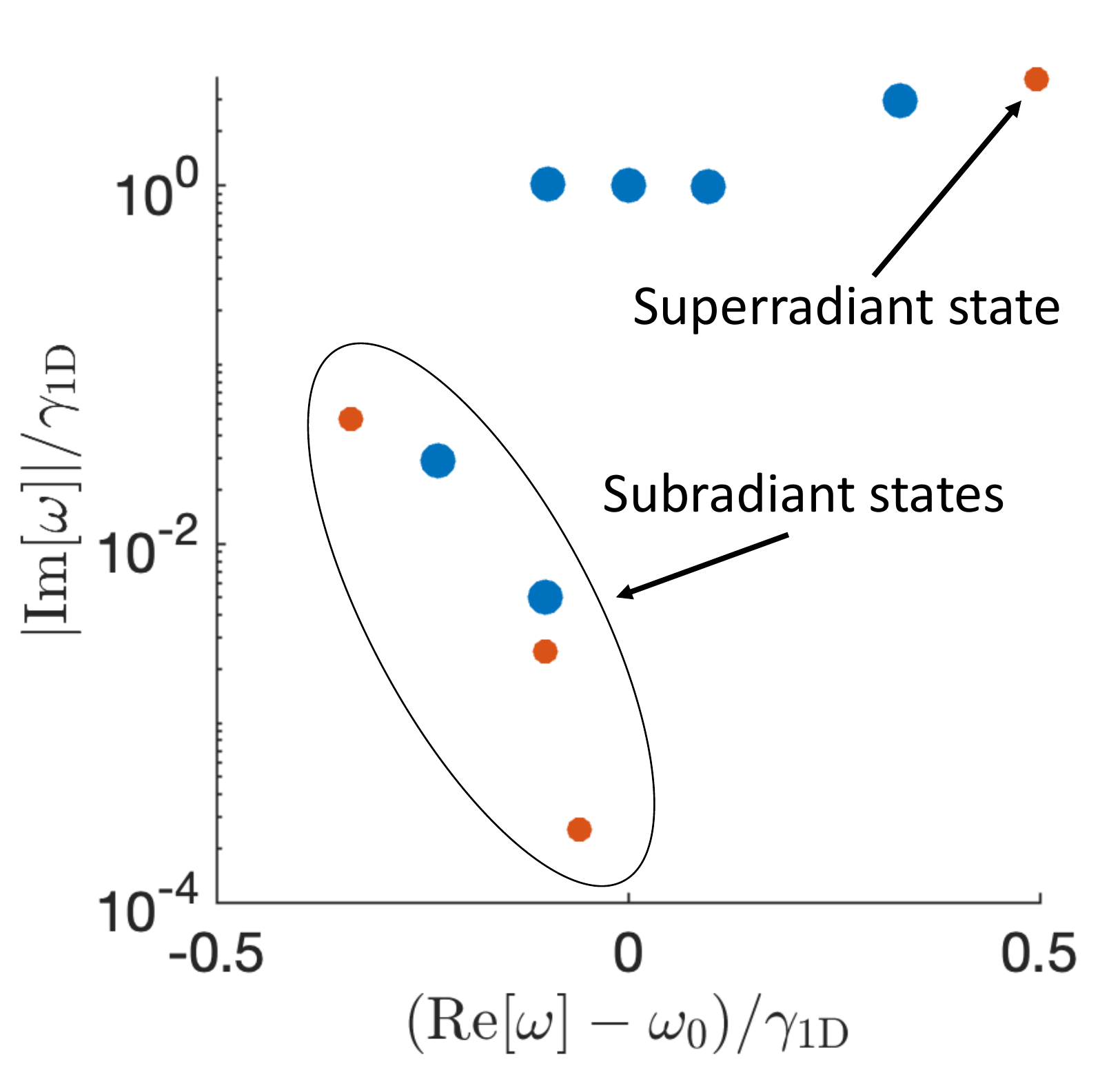}
\caption{Complex frequency spectrum for the following system parameters: number of qubits $N=4$, period of the system $\varphi=0.1$. Orange dots denote single photon states, blue dots correspond to the two-photon states.} \label{fig:3}
\end{figure}

For reference, we also present in  Fig.~\ref{fig:3} the complex spectrum of the system eigenfrequencies, calculated for the same system parameters as in Fig.~\ref{fig:2}.
Orange dots correspond to the single-excited states. They have been obtained as eigenvalues of the  effective Hamiltonian matrix $H_{mn}$, defined in Eq.~\eqref{eq:G}. The brightest state, with the largest imaginary  part, corresponds to the superradiant state with the decay rate $\approx N\gamma_{\rm 1D}$. Three other dots correspond to the single-excited subradiant states.
Blue dots show the spectrum of double-excited states. It  has been calculated following Ref.~\cite{Ke2019}. The eigenfrequencies were found by diagonalizing Eq.~\eqref{eq:Sh2}, given in the Appendix. The calculation demonstrates that there exists one superradiant mode, two subradiant ones and also three modes with decay rates on the order of $\gamma_{\rm 1D}$. As discussed in Ref.~\cite{Ke2019}, these three eigenstates could be understood as the ``twilight'' states, which are a product of  the wavefunction with one photon being in the bright state and the other one being in the subradiant one.

Our calculation approach, outlined in detail in the Appendix~\ref{sec:Appendix}, allows us to evaluate the contributions from various single- and double-excited eigenstates into the total transmitted wavefunction $\psi(t_1,t_2)$ separately. Generally, the single-excited states manifest themselves in the dependence on $t_1$ and $t_2$, that is along the edges of the color map in Fig.~\ref{fig:2}(a). 
The double-excited states correspond to the dependences on $t_1\pm t_2$, that is diagonal and antidiagonal in Fig.~\ref{fig:2}(a). Hence, the role of different contributions can be singled out by examining the cross sections in the corresponding directions, shown in Fig.~\ref{fig:2}(b,d). Our analysis of contributions  of various super- and sub-radiant eigenstates and the directions, along which these contributions are manifested, is schematically summarized in Fig.~\ref{fig:2}(a). We will now discuss it in more detail.

Single- and double- excited subradiant states manifest themselves as the long-living tails in the wavefunction along the edges of the calculation domain in Fig.~\ref{fig:2}(a) and along its main diagonal, respectively. Black solid and dark red curves show the two corresponding cuts in Fig.~\ref{fig:2}(b). In order to distinguish between single- and double-excited subradiant states, we have  performed  calculations along the same cuts  that neglect all single-excited subradiant states and include just a superradiant single-excited mode [dotted curves in Fig.~\ref{fig:2}(b)].  Such approximation  well describes the initial fast decay of the wave functions for both curves and the tails along the main diagonal ($t_1=t_2$, red dotted curve). Thus, the tails along the main diagonal can be attributed to the double-excited subradiant states. On the other hand, this approximation significantly underestimates the values of the tails of the wavefunction for fixed $t_2=0.2$, as can be seen by the comparison of  solid black and dotted gray curves in Fig.~\ref{fig:2}(b). This indicates that the tails in the solid black curve are due to the single-excited subradiant states.

The single-particle superradiant state manifests itself on the anti-diagonal in the time domain. It should be then measured as a function of the time difference between two photons. This can be seen by comparing  the  solid green  curve in Fig.~\ref{fig:2}(d), calculated accounting for all single-particle states, with the dashed orange one, that includes only superradiant single-particle states. Such a single superradiant mode approximation correctly describes the shape of the central peak in the full calculation. We have also checked that in order to correctly describe the amplitude of this sharp central feature it is necessary to include all the double-excited states.

\subsection{Duration of the transmitted pulse}\label{sec:N}

\begin{figure}[t] 
\includegraphics[width=0.48\textwidth]{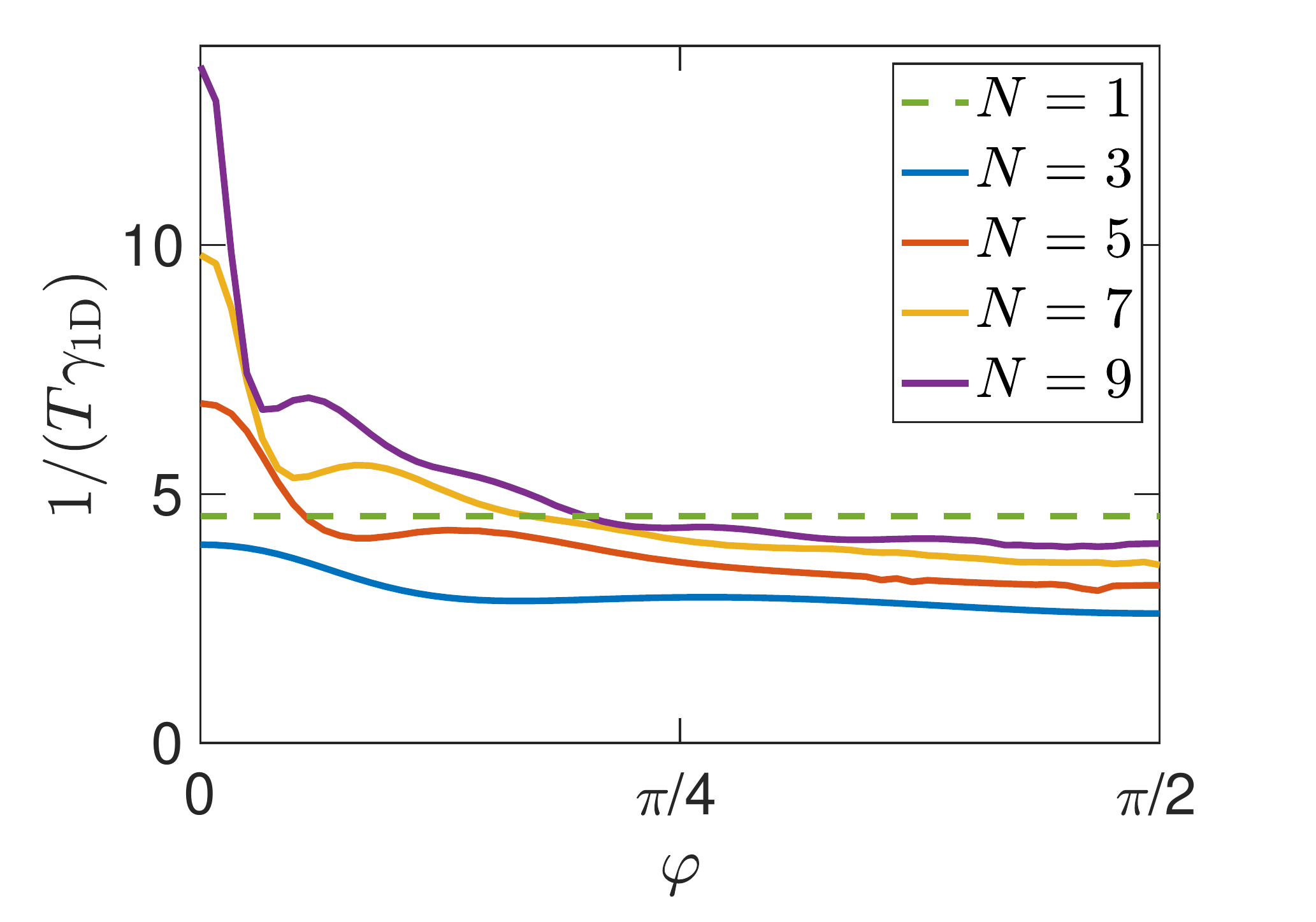}
\caption{Dependence of the inverse duration of the transmitted pulse $1/T$ on the period of the system $\varphi$ for  fixed numbers of qubits $N$.
} \label{fig:4}
\end{figure}

As the measure of the efficiency of the dark states' excitation, we introduce the quantity 
\begin{equation} \label{TPDur}
    T=\frac{\iint dt_1 dt_2|\psi^{\text{out}}_{t_1,t_2}|^2t_1}{\iint dt_1 dt_2|\psi^{\text{out}}_{t_1,t_2}|^2}
\end{equation}
the same way as it was done in Ref.~\cite{Poshakinskiy2012}. Taking into account the bosonic statistics of photons, we left only the time of the one photon $t_1$ under the integral (in general, we should look at the average times of both particles). The quantity $T$, by its definition, has the meaning of the duration of the transmitted pulse. Therefore, dark states' excitation efficiency is proportional to $T$. Fig.~\ref{fig:4} represents the dependence of the inverse duration of the transmitted pulse $1/T$ on the number of qubits $N$ and the period of the system $\varphi$. If all the qubits are located at one point ($\varphi=0$) then a short propagating pulse excites a superradiant state. This case corresponds to the maximum values of $1/T$ for each N in Fig.~\ref{fig:4}. If the qubits are located periodically at a distance $\varphi$ equal to $\pi/2$ then we notice the excitation of dark states. The period $\pi/2$ corresponds to the minimum values of $1/T$ for each $N$ in Fig.~\ref{fig:4}.

The quantity $1/T$ increases with the number of qubits for a fixed period $\varphi$ as the decay rate of the superradiant is equal to $N\gamma_{\text{1D}}$. The duration of the transmitted pulse does not increase with the number of qubits and, accordingly, the total length of the system because we consider Markovian approximation. This approximation implies the infinite speed of light, so the increase of the physical length of the system plays no role in the considered regime  of parameters~\cite{Poshakinskiy2012}.
\section{Summary}\label{sec:summary}
To summarize, we have developed a general analytical theory for the scattering of two-photon pulses from an array of two-level atoms, coupled to the waveguide. 
The wavefunction of the scattered pulse  has been obtained by a convolution of the known two-photon scattering matrix in the frequency domain with the Fourier transform of the incident pulse. In the case of a pulse duration being much shorter than the spontaneous emission lifetime, we were able to obtain a general analytical result for the scattered signal. This analytical expression, while being relatively cumbersome, considerably simplifies an interpretation of the scattered signal. Namely, it becomes possible to understand the role of the qualitatively different contributions corresponding  to various single-excited and double-excited eigenstates of the arrays with different radiative lifetimes (subradiant and superradiant states).

We have also studied the dependence of the average time it takes the array to emit two photons when being excited resonantly on the array length and period. The emission time  becomes generally shorter for longer structures, which can be explained by the formation of  superradiant single-excited photon states.  The longest emission times correspond to the structures  with the anti-Bragg period, equal to the quarter of the wavelength of light at the atom resonance frequency $\lambda/4$. This is due to the suppression of the superradiant states for the anti-Bragg structures.

Our results indicate that the time-dependent spectroscopy of  photon transmission can be an interesting complementary tool to the frequency domain analysis. It would be also instructive to generalize the results for the more complicated time dependence and entanglement structure of the input pulse. While this problem has already been analyzed in literature~\cite{Yang2022,Calajo2022},
the general effect of the excitation spectrum of the array on the quantum pulse transmission is far from being completely understood. For example, it would be interesting to examine what happens with the quantum light transmission through the Bragg structures with the period of $\lambda/2$~\cite{Poshakinskiy2012}, that can demonstrate strongly non-Markovian physics~\cite{PoshakinskiyBorrmann}.

\appendix

\section{Application of S-matrix method for the calculation of a delta-pulse transmission}\label{sec:Appendix}
Here we describe how to consider scattering of a short input two-photon pulse of the shape Eq.~\eqref{eq:psiin}. Its Fourier transform \eqref{eq:Fourier} is given just by 
    $\psi_{\omega_1,\omega_2}^{\rm in}=1$.
    Next, we define the frequency-integrated scattering matrix
\begin{equation}\label{eq:S12}
\begin{gathered}
    \tilde{S}(\omega_1',\omega_2')\equiv \iint \frac{d\omega_1 d\omega_2 }{(2\pi)^2} S(\omega'_{1},\omega'_{2}\leftarrow\omega_{1},\omega_{2})\\    =2t_{\omega'_1}t_{\omega'_2}+2\gamma_{\text{1D}}^2\sum_{i,j}s_i^-(\omega_1')s_i^-(\omega_2')Q_{ij}\\
    \times\int\frac{d\omega_1}{2\pi}s_j^+(\omega_1)s_j^+(\omega_1'+\omega_2'-\omega_1)\:.
\end{gathered}    
\end{equation}
In order to further proceed with the frequency integration it is instructive to expand the coupling coefficients 
\begin{equation}\label{eq:sj}
    s_j^{\pm}(\omega)=\sum_{\nu}\frac{s_j^{{\pm},\nu}}{\omega_{\nu}-\omega}
\end{equation}
as a sum of resonances at the single-excited state eigenfrequencies $\omega_\nu$. These are given just by the eigenvalues of the effective Hamiltonian matrix $H_{mn}$, defined in Eq.~\eqref{eq:G}.
Given Eq.~\eqref{eq:sj}, the frequency integration in the last line  of Eq.~\eqref{eq:S12}
results in 
\begin{equation}
\begin{gathered}
f_j^+(\omega_1'+\omega_2')\equiv\int\frac{d\omega_1}{2\pi}s_j^+(\omega_1)s_j^+(\omega_1'+\omega_2'-\omega_1)\\
=-\sum_{\mu,\nu}\int\frac{d\omega_1}{2\pi}\frac{s_j^{+,\nu}s_j^{+,\mu}}{(\omega_1-\omega_{\nu})(\omega_1+\omega_{\mu}-\omega_1'-\omega_2')}\\   =\mathrm{i}\sum_{\mu,\nu}\frac{s_j^{+,\nu}s_j^{+,\mu}}{(\omega_{\nu}+\omega_{\mu}-\omega_1'-\omega_2')}\:.
\end{gathered}\label{eq:f}
\end{equation}
In order to further proceed with the integration it is necessary to also expand the matrix $Q$ over the resonant terms. The resonances correspond to the double-excited states, found from the effective Hamiltonian
\begin{equation}\label{eq:Sh2}
\sum\limits_{m'n'=1}^N(\mathcal H+\mathcal U)_{mn,m'n'}\psi_{m'n'}=2\eps \psi_{mn}\:,
\end{equation}
with
$\mathcal H_{mn;m'n'}=\delta_{mm'}H_{nn'}+\delta_{nn'}H_{mm'}$
and 
$\mathcal U_{mn,m'n'}=\delta_{mn}\delta_{mm'}\delta_{nn'} U\:,$
where $m$ and $n$ are the coordinates of first and second excitation. Here, the coefficient $U$ describes  the anharmonicity of the qubit potential. In the considered case of two-level qubit, the limit $U\to \infty$ should be taken. Then, Eq.~\eqref{eq:Sigma} can be further simplified to
\begin{equation}
Q_{mn}=2(\rmi \eps-\gamma_{\rm 1D})\delta_{mn}+\sum\limits_{\nu=1}^{N(N-1)/2}\frac{2\rmi d^\nu_{m}d^{\nu}_{n}}{\eps_\nu-\eps}\label{eq:Q3}\:,
\end{equation}
where $\eps_\nu$ are the two-photon state energies found  from Eq.~\eqref{eq:Sh2},
and $d_m^\nu=\mathcal H_{mm;m'n'}\psi^\nu_{m'n'}$ with the normalization condition for two-photon states being $\sum_{m'n'}(\psi^\nu_{m'n'})^2=1$.

Using the expansions Eq.~\eqref{eq:Q3} and \eqref{eq:f} the scattering matrix
\eqref{eq:S12} becomes 
\begin{equation}\label{eq:S12b}
    \begin{gathered}
        \tilde{S}(\omega_1',\omega_2')=2t_{\omega_1'}t_{\omega_2'}\\
        +2\gamma_{\rm 1D}^2\sum_i s_i^-(\omega_1')s_i^-(\omega_2')u_i(\omega_1'+\omega_2')\:,
    \end{gathered}
\end{equation}
where
\begin{equation}
    \begin{gathered}
        u_i(\varepsilon)=-2\gamma_{\rm 1D}^2(\varepsilon+\mathrm{i}\gamma_{\rm 1D})\sum_{\nu,\mu}\frac{s_i^{+,\nu}s_i^{+,\mu}}{(\omega_{\nu}+\omega_{\mu}-2\varepsilon)}\\
        -2\gamma_{\rm 1D}^2\sum_{\kappa=1}^{N(N-1)/2}\frac{d_i^{\kappa}}{\varepsilon_{\kappa}-\varepsilon}\sum_{\nu,\mu}\frac{1}{\omega_{\nu}+\omega_{\mu}-2\varepsilon}\\
        \times\sum_{j=1}^N d_j^{\kappa}s_j^{+,\nu}s_j^{+,\mu}\:.
    \end{gathered}
\end{equation}
In order to simplify the notation it is convenient to relabel the indices so that $r=(\mu,\nu)$  and rewrite the same equation in a more general form:
\begin{equation}    u_i(\varepsilon)=\sum_r\frac{\mathrm{i}\varepsilon-\gamma_{\rm 1D}}{(\varepsilon_r-\varepsilon)}U_i^r+\sum_{rs}\frac{V_i^{rs}}{(\varepsilon_r-\varepsilon)(\varepsilon_s-\varepsilon)}\:,
\end{equation}
where 
\begin{equation}
    U_i^r=\mathrm{i}s_i^{+,\nu}s_i^{+,\mu}
\end{equation}
and
\begin{equation}
    V_i^{rs}=(-1)d_i^s\left(\sum_{j=1}^N d_j^s s_j^{+,\nu}s_j^{+,\mu}\right)\:.
\end{equation}
We are now in position to substitute Eq.~\eqref{eq:S12b} into Eq.~\eqref{eq:psit} and integrate over frequiencies $\omega_{1,2}$    to find the output wave function in the form
\begin{equation}
\begin{gathered}
  \psi_{t_1,t_2}^{\text{out}} =\psi_{t_1,t_2}^{\text{out, coh}}+\psi_{t_1,t_2}^{\text{out, incoh}}\\
  =\frac{1}{2}\iint\frac{d\omega_1'd\omega_2'}{(2\pi)^2}\tilde{S}(\omega_1',\omega_2')e^{-\mathrm{i}\omega_1't_1}e^{-\mathrm{i}\omega_2't_2}\:.
\end{gathered}
\end{equation}
The coherent part of the output wave function is given by the following expression
\begin{equation}
\psi_{t_1,t_2}^{\text{out, coh}}=y(t_1)y(t_2) \:,\label{PsiCoh}
\end{equation}
with
\begin{equation}\label{eq:psiout1}
    y(t_1)=\int\frac{d\omega_1'}{2\pi}t(\omega_1')e^{-\mathrm{i}\omega_1't_1}=\mathrm{i}\theta(t_1)\sum_{\mu}e^{-\mathrm{i}\omega_{\mu}t_1}t_{\mu}\:,
\end{equation}
\begin{equation}\label{eq:psiout2}
    y(t_2)=\int\frac{d\omega_2'}{2\pi}t(\omega_2')e^{-\mathrm{i}\omega_2't_2}=\mathrm{i}\theta(t_2)\sum_{\nu}e^{-\mathrm{i}\omega_{\nu}t_2}t_{\nu}\:.
\end{equation}
The incoherent part of the output wave function is written as
\begin{equation}
\begin{gathered}
\psi_{t_1,t_2}^{\text{out, incoh}}=\gamma_{\rm 1D}^2 \sum_{i,\mu,\nu}s_i^{-,\nu}s_i^{-,\mu}\bigg[\sum_r U_i^rL_{\nu \mu r}(t_1,t_2)\\
+\sum_{rs}V_i^{rs}M_{\nu \mu rs}(t_1,t_2)\bigg] \label{PsiIncoh}
\end{gathered}
\end{equation}
where
\begin{equation}\label{eq:psiout3}
\begin{gathered}
    L_{\nu \mu r}(t_1,t_2)=\theta(t_1)\theta(t_2)\iint \frac{d\omega_1'd\omega_2'}{(2\pi)^2}e^{-\mathrm{i}\omega_1't_1}\\
    \times e^{-\mathrm{i}\omega_2't_2}
    \frac{(\mathrm{i}\varepsilon-\gamma_{\rm 1D})}{(\varepsilon_r-\varepsilon)}\frac{1}{(\omega_{\nu}-\omega_1')(\omega_{\mu}-\omega_2')}\:,
\end{gathered}
\end{equation}
and
\begin{equation}
\begin{gathered}\label{eq:psiout4}
    M_{\nu\mu r s}(t_1,t_2)    =\theta(t_1)\theta(t_2)\iint\frac{d\omega_1'd\omega_2'}{(2\pi)^2}e^{-\mathrm{i}\omega_1't_1}\\
    \times e^{-\mathrm{i}\omega_2't_2}
    \frac{1}{(\varepsilon_r-\varepsilon)(\varepsilon_s-\varepsilon)} \frac{1}{(\omega_{\nu}-\omega_1')(\omega_{\mu}-\omega_2')}\:.
\end{gathered}
\end{equation}
For Eq.~\eqref{eq:psiout1} and Eq.~\eqref{eq:psiout2} we used the expansion of transmission coefficient over one particle resonant terms
\begin{equation}
    t(\omega)=\sum_{\mu}\frac{t_{\mu}}{\omega-\omega_{\mu}}.
\end{equation}
The integrals in Eq.~\eqref{eq:psiout3}, Eq.~\eqref{eq:psiout4} are readily found by Cauchy theorem e.g. in Mathematica. As a result, we obtain the following expressions
\begin{equation}
    \begin{gathered}
       L_{\nu \mu r}(t_1,t_2)=\theta(t_1)\theta(t_2)\Big[(\mathrm{i}\varepsilon_r-1)\\
       \times\Big(\theta(t_1-t_2)
       e^{-\mathrm{i}\omega_{\nu}(t_1-t_2)-2\mathrm{i}\varepsilon_rt_2}\\
       +\theta(t_2-t_1)e^{-\mathrm{i}\omega_{\mu}(t_2-t_1)-2\mathrm{i}\varepsilon_rt_1}\Big)\\
       -(\mathrm{i}(\omega_{\mu}+\omega_{\nu})/2-\gamma_{\rm 1D})e^{-\mathrm{i}\omega_{\nu}t_1-\mathrm{i}\omega_{\mu}t_2}\Big]\\
       \times\frac{1}{\varepsilon_r-(\omega_{\nu}+\omega_{\mu})/2},
    \end{gathered}
\end{equation}

\begin{equation}
\begin{gathered}
    M_{\nu\mu r s}(t_1,t_2)=\theta(t_1)\theta(t_2)\theta(t_2-t_1)\\
    \Big[(\varepsilon_r-(\omega_{\mu}+\omega_{\nu})/2)e^{-\mathrm{i}\omega_{\mu}(t_2-t_1)-2\mathrm{i}\varepsilon_st_1}\\
    -(\varepsilon_s-(\omega_{\mu}+\omega_{\nu})/2)e^{-\mathrm{i}\omega_{\mu}(t_2-t_1)-2\mathrm{i}\varepsilon_r t_1}\\
    -(\varepsilon_r-\varepsilon_s)e^{-\mathrm{i}\omega_{\nu}t_1-\mathrm{i}\omega_{\mu}t_2}\Big]\\
    \times\frac{1}{(\varepsilon_r-(\omega_{\nu}+\omega_{\mu})/2)(\varepsilon_s-(\omega_{\nu}+\omega_{\mu})/2)(\varepsilon_r-\varepsilon_s)}\\
    +(t_1 \leftrightarrow t_2, \omega_{\nu} \leftrightarrow \omega_{\mu}).
\end{gathered}
\end{equation}

\bibliography{twophotonpulse}
\end{document}